\documentclass[aps,preprint,onecolumn]{revtex4}%
\usepackage{amssymb}
\usepackage{amsmath}
\usepackage{amsfonts}
\usepackage{graphicx}%
\providecommand{\U}[1]{\protect\rule{.1in}{.1in}}
\begin{document}
\title{Ferromagnetism in the Blume-Emery-Griffiths model on finite-size Cayley tree}
\author{Wen-Jun Chen}
\author{Xiang-Mu Kong}
\thanks{Corresponding author}
\email{kongxm@mail.qfnu.edu.cn (X.-M. Kong)}
\affiliation{Shandong Provincial Key Laboratory of Laser Polarization and Information
Technology, Department of Physics, Qufu Normal University, Qufu 273165, China}
\keywords{}
\pacs{03.67.Mn, 03.65.Ud, 75.10.Jm}

\begin{abstract}
The ferromagnetic properties of the spin-1 BEG model on finite-size Cayley
tree are investigated using the exact recursion method. The spontaneous
magnetization of the system is studied in detail for different values of the
reduced crystal-field interaction $D/J$, and it is found that there is an
unusual\ behavior (anti-Curie temperature) when $D/J>2.0$. We also obtain the
Curie temperature of this finite-size system. When the system size is large
enough, our results will fit well with that in the thermodynamic limit.

\end{abstract}
\volumeyear{year}
\volumenumber{number}
\issuenumber{number}
\eid{identifier}
\date[Date text]{date}
\received[Received text]{date}

\revised[Revised text]{date}

\accepted[Accepted text]{date}

\published[Published text]{date}

\startpage{1}
\endpage{ }
\maketitle

\section{Introduction}

The Cayley tree \cite{Cayley tree} which is also called Bethe lattice
\cite{Bethe} was firstly investigated by Kurata \textit{et al. }\cite{Kurata}.
Then Domb studied the Ising model on such lattice and demonstrated that a
Bethe-Peierls approximation is exact for the Bethe lattice
\cite{Domb,Domb1984}. Over the years, the thermodynamic properties of
the\ Ising system on this lattice have been extensively investigated
\cite{Eggarter,Heimburg,Tanaka,Katsura}. As an expanded Ising model, the
Blume-Emery-Griffiths (BEG) model \cite{BEG,Griffiths}, which is characterized
by bilinear and biquadratic exchange interactions and crystal-field
interaction \cite{Blume,Capel}, has played an important role in the
development of the theory of tricritical phenomena. This model has been
studied by a variety of techniques, e.g., the generalized Bethe-Peierls
approximation \cite{Obokata and Oguchi,Iwashita,Uryu}, the\ effective-field
theory \cite{Chakraborty,Chakraborty2,Fittipaldi}, the generalized
constant-coupling approximation theory with two parameters
\cite{Takahashi1979,Takahashi1980}, the exact recursion relations method
\cite{C.Ekiz,C.Ekiz1,C.Ekiz2} and so on.

All of these systems mentioned above are studied in the case of the
thermodynamic limit. As we all know that the systems studied by the methods of
experiment and numerical simulation are all finite. So the research on the
finite-size system is much more meaningful. It was not until recently that a
new exact expression of finite-size system for the zero-field magnetization
was established \cite{Melin1996}. Then the corresponding exact expression in
closed form for the zero-field susceptibility was given by T. Stosic
\textit{et al. }\cite{B.D.Stosic,T.Stosic,I.P.Fittipaldi}. To our knowledge no
exact calculation has been made in the field regarding the properties of the
spin-1 BEG model on the finite-size Cayley tree yet.

In this paper we investigate the effect of the finite-size system on
the\ ferromagnetic properties in detail. The expression of the magnetization
for this system with different values of the reduced crystal-field
interaction\ is derived and the Curie temperature is obtained. The results are
compared with that of the case in the thermodynamic\ limit.

\section{Model and formulation\label{model}}

At the beginning, we give a brief description of the construction of the
Cayley tree. Starting from a single point $0$, the central one of the graph
\cite{R.J.Baxter}, we add $q$ different points connected to the central point
which may be called\ "the first shell". Then each point of the first shell is
joined to $q-1$ new points. So the points of the first shell have
$q(q-1)$\ nearest neighbors in total which form the second shell. The number
of shells is also defined as generation number $n$ while $q$ as coordination
number. If we continue in this way, the entire structure of the Cayley tree is formed.

The Hamiltonian of the BEG model on the Cayley tree is given by%
\begin{equation}
H=-J\sum_{\left\langle ij\right\rangle }S_{i}S_{j}-K\sum_{\left\langle
ij\right\rangle }S_{i}^{2}S_{i}^{2}+D\sum_{i}S_{i}^{2}-h\sum_{i}S_{i}\text{,}
\label{H}%
\end{equation}
where $S_{i}(=\pm1,0)$ is the spin at site $i$, the summation $\sum_{i}$ runs
over all the sites and $\sum_{\left\langle ij\right\rangle }$ denotes
summation over\ all the nearest-neighbor pairs.$\ J$, $K$, $D$ and $h$
describe the bilinear exchange, biquadratic interaction, crystal-field (or
single-ion anisotropy) interaction and external magnetic field, respectively.
This Hamiltonian was originally proposed to explain the phase separation and
superfluidity in $^{\text{3}}$He-$^{\text{4}}$He mixtures \cite{BEG}.

The partition function of the above system can be written as%
\begin{align}
Z  &  =\sum\exp\left(  -\beta H\right) \nonumber\\
&  =\sum_{S}\exp\left[  \beta\left(  J\sum_{\left\langle ij\right\rangle
}S_{i}S_{j}+K\sum_{\left\langle ij\right\rangle }S_{i}^{2}S_{i}^{2}-D\sum
_{i}S_{i}^{2}+h\sum_{i}S_{i}\right)  \right]  \text{,} \label{Z}%
\end{align}
where $\beta=1/k_{B}T$, $k_{B}$ is Boltzmann constant and $T$ is the absolute
temperature. The summation $\sum_{S}$ in Eq. (\ref{Z}) goes over all spin
configurations of the system.

Without loss of generality, we consider a Cayley tree of\textit{ }$n$
generations with branch number $B=5$ (coordination number minus one). Then,
the $n$-generation branch consists of $N_{n}=\frac{5^{n+1}-1}{4}$ spins, while
the 0-generation being a single spin. Let $Z_{n}^{(+)}$, $Z_{n}^{(-)}$ and
$Z_{n}^{(0)}$ be the partial partition functions of the system, with the
central spin takes values +1, 0 and -1 respectively. Based on Eq. (\ref{Z}),
we can obtain the recursion relations \cite{Eggarter} as%
\begin{equation}
Z_{n+1}^{\left(  \pm\right)  }=e^{-\beta D\pm\beta h}\left(  e^{\pm\beta
J+\beta K}Z_{n}^{\left(  +\right)  }+Z_{n}^{\left(  0\right)  }+e^{\mp\beta
J+\beta K}Z_{n}^{\left(  -\right)  }\right)  ^{5} \label{Z+-}%
\end{equation}
and%
\begin{equation}
Z_{n+1}^{\left(  0\right)  }=\left(  Z_{n}^{\left(  +\right)  }+Z_{n}^{\left(
0\right)  }+Z_{n}^{\left(  -\right)  }\right)  ^{5}\text{.} \label{Z0}%
\end{equation}

The partition function Eqs. (\ref{Z+-}) and (\ref{Z0}) can be differentiated
with respect to field, thus the recursion relations for the field derivatives
of the partition function are easily written as%
\begin{align*}
\frac{\partial Z_{n+1}^{\left(  \pm\right)  }}{\partial\beta h} &  =\pm
e^{^{-\beta D\pm\beta h}}\left(  e^{\pm\beta J+\beta K}Z_{n}^{\left(
+\right)  }+Z_{n}^{\left(  0\right)  }+e^{\mp\beta J+\beta K}Z_{n}^{\left(
-\right)  }\right)  ^{5}\\
&  +5e^{^{-\beta D\pm\beta h}}\left(  e^{\pm\beta J+\beta K}Z_{n}^{\left(
+\right)  }+Z_{n}^{\left(  0\right)  }+e^{\mp\beta J+\beta K}Z_{n}^{\left(
-\right)  }\right)  ^{4}\\
&  \times\left(  e^{\pm\beta J+\beta K}\frac{\partial Z_{n}^{\left(  +\right)
}}{\partial\beta h}+\frac{\partial Z_{n}^{\left(  0\right)  }}{\partial\beta
h}+e^{\mp\beta J+\beta K}\frac{\partial Z_{n}^{\left(  -\right)  }}%
{\partial\beta h}\right)  \text{,}%
\end{align*}%
\begin{equation}
\frac{\partial Z_{n+1}^{\left(  0\right)  }}{\partial\beta h}=5\left(
Z_{n}^{\left(  +\right)  }+Z_{n}^{\left(  0\right)  }+Z_{n}^{\left(  -\right)
}\right)  ^{4}\left(  \frac{\partial Z_{n}^{\left(  +\right)  }}{\partial\beta
h}+\frac{\partial Z_{n}^{\left(  0\right)  }}{\partial\beta h}+\frac{\partial
Z_{n}^{\left(  -\right)  }}{\partial\beta h}\right)  .\label{1}%
\end{equation}

\bigskip Starting from a single spin (0-th generation branch) we have%
\[
Z_{0}^{\left(  \pm\right)  }=e^{^{-\beta D\pm\beta h}}\text{, \ \ }%
Z_{0}^{\left(  0\right)  }=1\text{,}%
\]
and%
\[
\frac{\partial Z_{0}^{\left(  \pm\right)  }}{\partial\beta h}=\pm e^{^{-\beta
D\pm\beta h}}\text{, \ \ }\frac{\partial Z_{0}^{\left(  0\right)  }}%
{\partial\beta h}=0\text{.}%
\]

Then the magnetization of a site in the Cayley tree can be written as%
\begin{equation}
\left\langle m\right\rangle _{n}^{\pm}=\frac{1}{N_{n}}\frac{1}{Z_{n}^{\left(
\pm\right)  }}\frac{\partial Z_{n}^{\left(  \pm\right)  }}{\partial\beta
h}\text{.} \label{M}%
\end{equation}

The exact recursion relations\ and the magnetization\ expression\ allow us to
study the thermodynamic behavior of the system in detail. In next section, we
will give the numerical results of the magnetization.

\section{Numerical results and discussions\label{results}}

\ In the following, we investigate numerically the magnetization of the BEG
model on this finite-size Cayley tree. Based on the above Eqs.~(\ref{H}%
-\ref{M}), taking the limit $h\rightarrow0$, we\ make a detailed calculation
to the magnetization for various strengths of the interaction $J$, $K$ and $D$
and for several system sizes $n$.

The magnetization as a function of temperature is shown in Fig. 1 for various
reduced values of crystal-field interaction $D/J$ while $K/J=1$, $n=12$. The
curve corresponding to $D/J=2.0$ separates the curves into two different
trend. In the case of\ $D/J>2.0$, there is an anti-Curie temperature $T_{aC}$
\cite{Capel}, i.e., a temperature below which the magnetization vanishes, in
addition to the Curie temperature $T_{C}$. As $D/J$ decreases, the anti-Curie
temperature drops to lower values and the maximum magnetization becomes
larger. We can see that the anti-Curie temperature vanishes in the case of
$D/J=2.0$ and the magnetization decreases steadily from its saturation value
$0.58$ to zero with growing temperature. In the case of $D/J<2.0$, the maximum
magnetization of curves increases and it approaches to unity at zero
temperature for $D/J=1.5$. Moreover the magnetization curve would keep the
conventional shape for even smaller $D/J$ values that is not shown in this
figure. An approximative value of the deduced Curie temperature $T_{C}=1.4$
can also be obtained for this two-order transition.

In Fig.~2 we present the magnetization as a function of temperature for
several system sizes $n=3,6,9,12$ and different\ values of $D/J$. It is seen
that the curves exhibit a slow\ decay\ of magnetic ordering with the increase
of the system sites. The larger the system size, the sharper the magnetization
curve is. If the system sizes could get large enough value, the magnetization
curve of finite-size system would become close to that in thermodynamic
limit.\ The unusual shape of the figure is due to the complicated interactions
including the bilinear exchange and biquadratic exchange. This result is in
agreement with the system in thermodynamic limit which is studied by K. G.
Chakraborty\emph{ }\textit{et al. }\cite{J.W.Tucker,K.G.Chakraborty}.

\section{Conclusions\label{conclusions}}

In this paper, using the exact recursion relations, we present an exact
calculation for the spontaneous magnetization of the spin-1 BEG system on the
finite-size Cayley tree. The magnetization properties are studied in detail
for different values of the crystal-field interaction and system size of the
Cayley tree. It is shown that the magnetization curves exhibit some unusual
features including an anti-Curie temperature as the variance of the strength
of the crystal-field interaction. The Curie temperature is also obtained. The
curves would become close to that in the thermodynamic limit with the increase
of the system size.

\section{Acknowledgments}

This work was supported by the National Natural Science foundation of China
under Grant NO. 10775088, and the Shandong Natural Science foundation under
Grant NO. Y2006A05. One of the authors (Chen) thanks Shuxia Chen, Sai Wang and
Shengxin Liu for fruitful discussions.

FIG.~1: The magnetization $M$ as a function of $T$ for $K/J=1,n=12$ and $B=5.$
The successive curves from (a) to (b) are for $D/J=2.3,2.1,2.0,1.9,1.5$ and
$1$ respectively. There exhibits a two-order phase transition.

FIG. 2: The magnetization versus temperature $T$ for different $D/J$ and
$n=3,6.9,12$. The successive curves from (a) to (f) are for
$D/J=2.3,2.1,2.0,1.9,1.5$ and $1$ respectively. The slope of each subgraph
curve steepens with the increase of the system sizes $n.$

\newpage%
%TCIMACRO{\FRAME{ftbpFU}{3.3148in}{2.6325in}{0pt}{\Qcb{The magnetization $M$ as
%a function of $T$ for $K/J=1$, $n=12$ and $B=5.$ The successive curves from
%(a) to (b) are for $D/J=2.3,2.1,2.0,1.9,1.5$ and $1$ respectively. There
%exhibits a two-order phase transition.}}{}{fig1.eps}%
%maintain-aspect-ratio TRUE;  display "ICON";  valid_file "F";
%width 3.3148in;  height 2.6325in;  depth 0pt;  original-width 3.9669in;
%original-height 3.1453in;  cropleft "0";  croptop "1";  cropright "1";
%cropbottom "0";  filename 'fig1.eps';file-properties "XNPEU";}} }%
%BeginExpansion
\begin{figure}
[ptb]
\begin{center}
\includegraphics[
height=2.6325in,
width=3.3148in
]%
{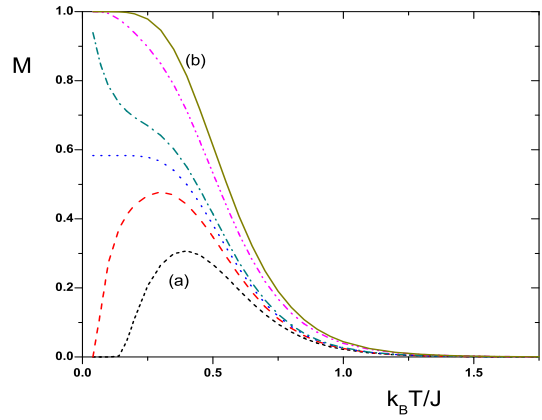}%
\caption{The magnetization $M$ as a function of $T$ for $K/J=1$, $n=12$ and
$B=5.$ The successive curves from (a) to (b) are for $D/J=2.3,2.1,2.0,1.9,1.5$
and $1$ respectively. There exhibits a two-order phase transition.}%
\end{center}
\end{figure}
%EndExpansion
%

%TCIMACRO{\FRAME{ftbpFU}{4.529in}{4.465in}{0pt}{\Qcb{The magnetization versus
%temperature $T$ for different $D/J$ and $n=3,6.9,12$. The successive curves
%from (a) to (f) are for $D/J=2.3,2.1,2.0,1.9,1.5$ and $1$ respectively. The
%slope of each subgraph curve steepens with the increase of the system sizes
%$n$.}}{}{fig2.eps}{\special{ language "Scientific Word";  type "GRAPHIC";
%display "ICON";  valid_file "F";  width 4.529in;  height 4.465in;  depth 0pt;
%original-width 2.9577in;  original-height 4.0075in;  cropleft "0";
%filename 'fig2.eps';file-properties "XNPEU";}} }%
%BeginExpansion
\begin{figure}
[ptb]
\begin{center}
\includegraphics[
height=4.465in,
width=4.529in
]%
{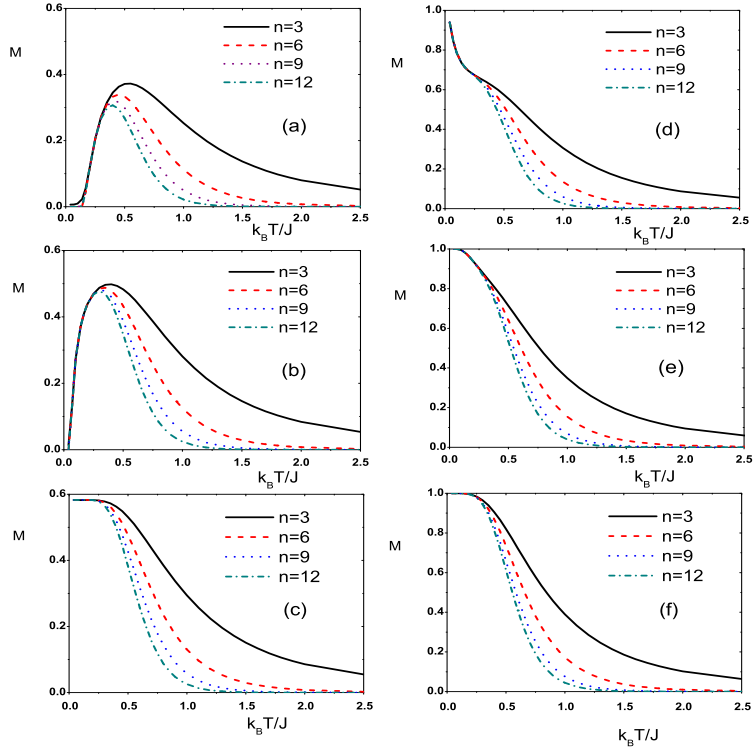}%
\caption{The magnetization versus temperature $T$ for different $D/J$ and
$n=3,6.9,12$. The successive curves from (a) to (f) are for
$D/J=2.3,2.1,2.0,1.9,1.5$ and $1$ respectively. The slope of each subgraph
curve steepens with the increase of the system sizes $n$.}%
\end{center}
\end{figure}
%EndExpansion

\bigskip

\end{document}